\documentclass[
 reprint,
 amsmath,amssymb,
 aps,
pra,
]{revtex4-2}

\usepackage{graphicx}% Include figure files
\usepackage{dcolumn}% Align table columns on decimal point
\usepackage{bm}% bold math
\usepackage[utf8]{inputenc}
\usepackage[T1]{fontenc}

\begin{document}

\preprint{APS/123-QED}

\title{Active Learning reduces academic risk of students with non-formal reasoning skills.\\ Evidence from an introductory physics massive course in a Chilean public university}

\author{Guillaume Lagubeau}
\email{guillaume.lagubeau@usach.cl}%Lines break automatically or can be forced with \\

\author{Silvia Tecpan}
\email{silvia.tecpan@usach.cl}
\author{Carla Hernandez}
\email{carla.hernandez.s@usach.cl}
 
\affiliation{%
 Universidad de Santiago de Chile, Departamento de Física
}%

\date{\today}% It is always \today, today,
             %  but any date may be explicitly specified

\begin{abstract}
We present the findings of a pilot plan of active learning implemented in introductory physics in a Chilean public university.
The model is research based as it considered a literature review for adequate selection and design of activities, consistent with the levels of students' reasoning skills.
The level of scientific reasoning is positively correlated to student success. 
By contrast to a control group of students following traditional lectures, we observed a significant reduction in failure rate for students that do not yet posses formal scientific reasoning. 
This profile of student being the majority, we conclude that implementing active learning is particularly suited to first year of higher education in the context of a developing country. 
It fits the particularities of student profile and typical classroom size, leading to learning improvement and reduction of academic risk as well as being financially sound.

\end{abstract}

%Use showkeys class option if keyword
                              %display desired
\maketitle

%\tableofcontents

\section{Introduction}

Low student enrollment and high attrition rates in Science, Technology, Engineering, and Mathematics (STEM) education are part of the major contemporary challenges in higher education \cite{sithole2017student}. 
As a consequence, introductory physics courses usually becomes filter courses for numerous engineering students \cite{vasquez2015early}. 
Indeed, this is reflected by our institution's historical approval rate in first semester introductory physics course taught for all engineering careers. 
In an effort to reduce academic risk, as well as to better prepare engineering students to 21st century, a methodological change in the teaching method \footnote{MEI usach 2014} \footnote{https://www.consortium2030.cl/} was decided in our institution.
Strong evidences accumulated over the past 35 years of significant gains in learning physics \cite{freeman2014active,crouch2001peer,sharma2010use} using active learning motivated a pilot program of implementation of research based \cite{bubou2017research} active learning in introductory physics course.

The pilot program main challenges were making it realistically scalable for encompassing a large enrollment course (1600 students) and adapting innovative strategies to the context of our traditional and public university. 
Our institution has a voluntarily inclusive access policy that favors admissions to higher education of students with very heterogeneous profiles.
Students originates from diverse ways of entry such as university selection test, and high school ranking, among others \cite{santelices2018high,santelices2019}. 
As deep understanding of physics concepts requires formal reasoning \cite{lawson2010teaching,fabby2015examining}, it is essential to characterize the level of scientific reasoning of our introductory physics course students before tailoring a teaching sequence consistent with their profiles \cite{coletta2007you, coletta2005interpreting}.

In this article we report on a professor training model implemented at piloting level to transform introductory physics courses for engineering programs. 
This model is inspired by previous similar experiences both in Latin America \cite{zavala2007innovative, zavala2008evaluation, auyuanet2018fisicactiva, hernandez2018correct} and others countries \cite{redish2014oersted, rudolph2014introduction, deslauriers2011improved}. 
By comparing the results of an experimental group following active learning and a control group, we found a 9.1\% reduction in failure rate (see table \ref{Table:FR_global}), statistically significant and coherent with previous reports \cite{freeman2014active}. 
In addition, students with transitional reasoning level, being the majority in our context, benefited most of the innovation.

\begin{table}[h]
\caption{\textbf{Failure rate in introductory physics.} $N$ is the universe of students and $p(H_O)$ is the probability that failure rate decreased by introducing active learning.}
\begin{ruledtabular}
\begin{tabular}{cccccc}
N & traditional & active 
 & variation & $p(H_0)$\\
\hline
\textbf{304} & \textbf{45.7\%} & \textbf{36.1\%} &\textbf{-9.1\%} & \textbf{93.7\%} \\
\end{tabular}
\end{ruledtabular}
\label{Table:FR_global}
\end{table}

\section{Theoretical Framework}
The level of scientific reasoning is well reported to be a determining factor in academic success in the first years of university science courses \cite{coletta2005interpreting, coletta2007you,fabby2015examining}. 
Methodologies that promote active learning, that is designing lectures where the student is intellectually active \cite{meltzer2012resource} have been found to improve scientific reasoning \cite{ding2013detecting,fabby2013relationship}.

For designing classroom activities, we used active learning strategies focused on the need to enhance conceptual learning, problem solving skills, collaborative work and hypothesis generation among other skills required for the training of engineers in the 21st century. 
In particular, we used tutorials \cite{mcdermott1997tutoriales}, interactive lecture demonstration \cite{sokoloff2004interactive}, peer instruction \cite{mazur1999peer}, sense-making tasks \cite{hieggelke2015tipers} and collaborative Solving Problems \cite{heller1992teaching} fitting the program of the course (introductory mechanics and statics).
In supplementary material, we provide a list of the activities used for the active learning sessions.

\begin{figure}[h]
\includegraphics[width=1\linewidth]{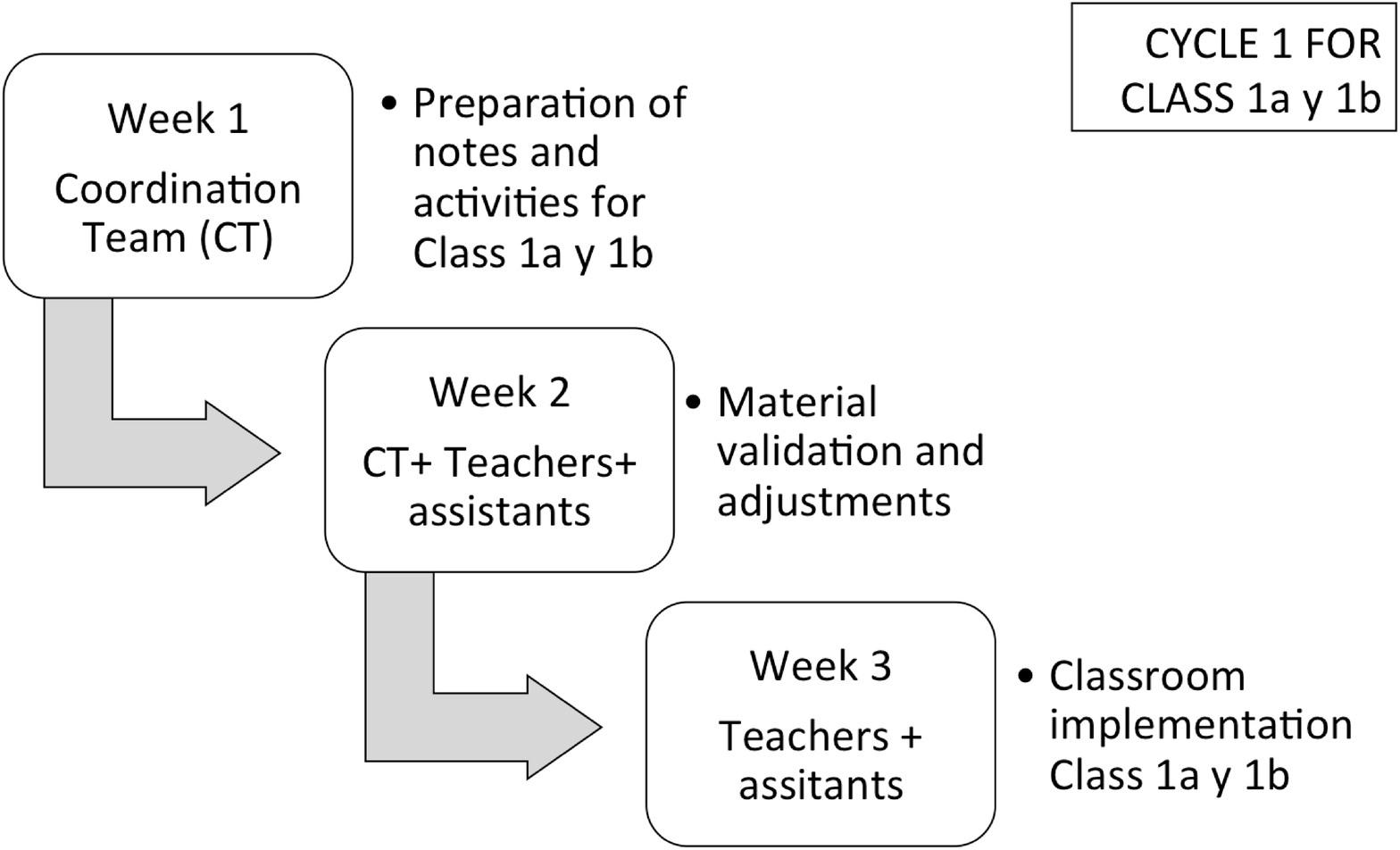}%

\includegraphics[width=0.75\linewidth]{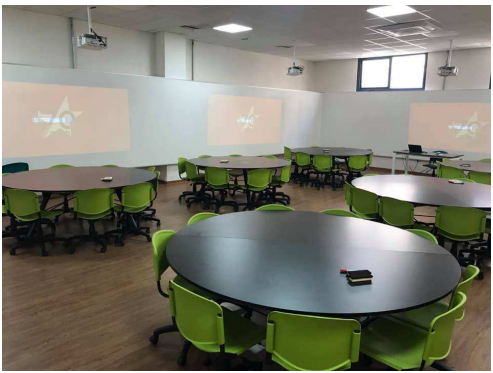}%
\caption{Top: work flow. Bottom: student centered classroom.}
\label{fig:sala}
\end{figure}

The teaching material was designed by a coordination team, distinct from the group of teachers. 
Interactive lectures were given by a professor accompanied by a teaching assistant. 
The professors had already taught the same course in traditional way previously, and undergraduate students of our university served as teaching assistants were. 
Both professors and teaching assistants were present simultaneously during lecture to support the student's learning process. 
Implementing active learning strategies in the classroom requires a preparation work that we structured in a three-stage cycle (fig.\ref{fig:sala} top).

The first stage consists in the coordination team reviewing the literature to prepare the activities that would be used in the two class occurring two weeks after. 
In the second stage, the proposed activities are reviewed and adjusted, if necessary in a meeting between the coordination team, the professors and the teaching assistants.
Finally, in stage three, teachers and assistants implement activities in the classroom. 
This process started two weeks before starting the semester and was repeated each week. 
%For sharing course content with professors and teacher  assistants, the coordinating team also created an online repository of material and the essential bibliographical references.

Classes are held in Scale-Up rooms \cite{beichner2007student}, build for this project, of capacity 54 students. 
Their design promotes student-centered learning as the teacher is not the focal point. 
Five projectors provide good visibility of projected material for all students.
The furniture consisted initially of normal tables and chairs, arranged in groups of 9 people to work in micro groups of 3 students, and latter large round desks that facilitate collaborative work were installed, as seen in figure \ref{fig:sala} bottom.
The entire wall of the room are covered with boards. In addition, blackboards of 60cm x 80cm are freely available in room for every group of students.
Finally, the room layout allowed an easy circulation of students and professors between desks.

Following the flipped classroom model \cite{fox2018learning,hernandez2017aula} material is provided to the students gradually: elements of theory and exercises to complement traditional classes, identical with those of the previous semester, were provided before the active session through a virtual institutional platform. 
Indeed this material is available to all the students, in active modality or not.

Such a methodological change may be associated to an increment in infrastructure and personal costs per student \cite{brewe2018costs}.
It is due to reducing sections sizes and thus increasing the number of sections in massive courses.
Due to the layout and infrastructure of the university, it is not the case in our university:
the average section of introductory physics is composed by 50 students (31 sections for approximately 1550 students in total). 
Scaling up the pilot plan would be realized at constant number of sections and professors. 
 
\section{Methods}
\subsection{\label{sec:Methods} Experimental design}
The pilot program concerned 4 sections from a total of 31 were experimental (active learning) and 4 sections of equivalent historical results as control group. 
Both groups reasoning skills were characterized by taking the Lawson Classroom Test of Scientific Reasoning \cite{Bao2018} at the start of the semester.

Both control and experimental groups followed the same weekly program of contents and had access to the same bibliography or online material.
Indeed, online material and evaluations were not innovated and were similar to previous semesters. 
The experimental group followed two active lectures of 1h30 each and an active exercise session of 1h30. 
The control group had the same schedule but using traditional teaching.
Three evaluations (identical for all students) were carried out with three problems each, prepared by a teaching committee not included in the implementation of the pilot plan.

\subsection{Failure rate analysis}
For consistency with the meta-analysis of Freeman \cite{freeman2014active}, we compare failure rate of students defined as the number of student failing the course divided by the total number of students. 
We only consider students that attended all the evaluations during the semester forming a universe of 304 students, 146 of which followed traditional learning and 158 active learning.
For determining the statistical significance of the difference in failure rate, we test the following hypothesis ($H_0$): ``introductory physics students following active lecture are less likely to fail than students following traditional lecture". 
The null hypothesis is then that ``introductory physics students following active lecture are as likely or more likely to fail than students following traditional lecture". 
Our statistical analysis is the following: failing or passing a course is a binary process. 
Thus, evaluating the failure rate is equivalent to estimate a so called ``cut efficiency". 
Bayes analysis allows to theoretically calculate the uncertainty of efficiency measurement due to size effect \cite{Paterno2004}. 
If $k$ is the number of positive cases and $n$ the population, the probability distribution of the efficiency is a beta distribution of parameters ($\alpha=k+1$, $\beta=n-k+1$). 
We can then calculate the density probability of the variation in failure rate (as shown in figure \ref{fig_global}) and therefore estimate p($H_0$).

\subsection{Scientific reasoning diagnostic}
During the two first weeks of the semester, students of the control and experimental groups took the Lawson Classroom Test of Scientific Reasoning \cite{Bao2018}. 
The test is composed of 24 questions, organised by pairs. 
Bao \cite{Bao2009} and Mashood \cite{Mashood2019} used the distribution of correct answer as an indicator of typical reasoning skills in first year of university, comparing different cultures. 
For our study, a total of 260 students took the reasoning test and were present in all evaluations of the semester (149 from active groups and 111 from control group).

\section{Results}
\subsection{Failure rate reduction}
 Failure rate was 45.2\% for students following traditional learning, comparable to the historic rate (45.7\%, considering only students that attended all evaluation). 
 By contrast, active learning students failing rate was 36.1\%, evidencing a 9.1\% improvement ($p(H_0)=0.937$) consistent with improvement reported in literature for physics and STEM \cite{freeman2014active}. 
 We conclude that active learning methodology is particularly well suited in the Chilean context for teaching introductory physics contents. 
 
 Separating results by gender, one can note that while in both modalities, female failure rate was higher than male failure rate, female students benefited more of the innovation than male students (see table \ref{Gender}): female students following traditional learning were 1.41 time more likely to fail than those following active learning (1.29 for male students).
 
 \begin{table}[h]

\caption{\textbf{Failure rate in introductory physics as a function of gender}. $p(H_O)$ is the probability that
that failure rate decreased by introducing active learning. Bold letters are used when $H_O$ is statistically likely.}
\begin{ruledtabular}
\label{tableSexo}
\begin{tabular}{cccccc}
& N & traditional & active 
 & variation & $p(H_0)$\\
\hline
\textbf{Female} & \textbf{76} & \textbf{67.9\%} & \textbf{47.9\%} &\textbf{-20.0\%} & \textbf{94.7\%} \\

\textbf{Male}& \textbf{228} & \textbf{39.8\%} & \textbf{30.9\%} & \textbf{-8.9\%}
& \textbf{90.7\%} \\

\end{tabular}
\end{ruledtabular}
 \label{Gender}
\end{table}
 
\subsection{Correlation between reasoning skills and active learning efficiency}

\begin{figure}[b]
\includegraphics[width=0.8\linewidth]{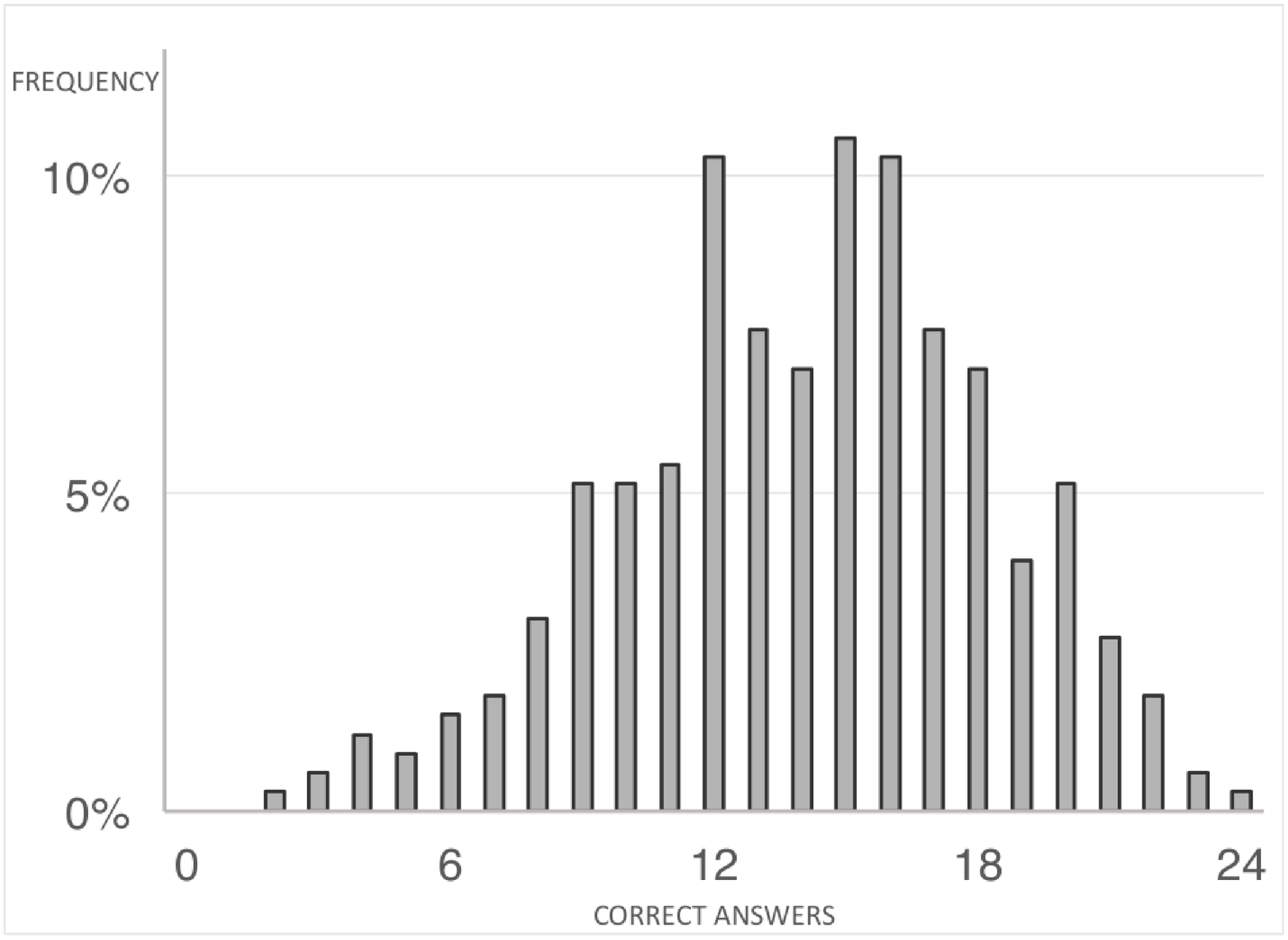}%

\includegraphics[width=0.8\linewidth]{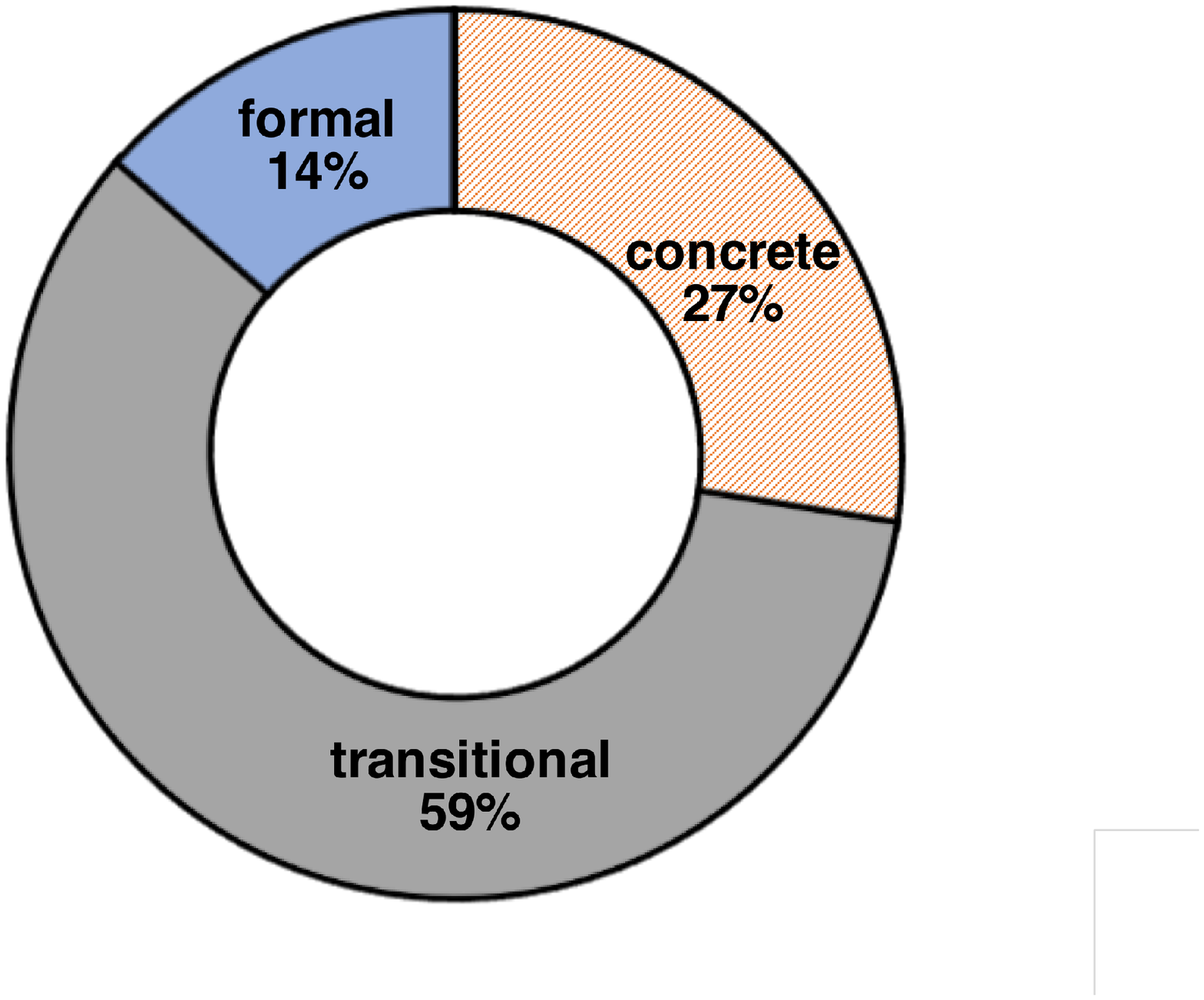}%
\caption{Results of the Lawson Classroom Test of Scientific Reasoning for introductory physics students, experimental and control groups combined (N=260). Top: distribution of students as a function of the number of correct answers}
\label{fig:Lawson}
\end{figure}

For comparison with previous cross cultural studies \cite{Bao2009, Mashood2019}, in fig.\ref{fig:Lawson} top, we present the distribution of students as a function of the number of correct answer to the Lawson Classroom Test of Scientific Reasoning. 
The average reasoning level in our students universe was found significantly lower than those reported in USA, China and India (see table \ref{Table_Lawson}) thus evidencing our local need to adapt to a profile of student with reasoning skills not yet fully developed. 
A pair analysis of answers allows to sorts the students in three reasoning level \cite{Bao2018}: the so called "concrete", "transitional" and "formal" reasoning skills. 
At the concrete level (0-4 pairs) students are able to classify objects and understand conservation, but not yet able to form hypotheses. 
At  the formal level (9-12 pairs), students can think abstractly and are able to control and isolate variables, among other similar tasks. 
At the transitional level (5-8 pairs), students are only capable of partial formal reasoning \cite{coletta2005interpreting}

In Fig.\ref{fig:Lawson} bottom we present the categorization of our introductory physics students. 
The majority of students are observed to be still in a transitional level of reasoning. 
In light of this diagnostic, obtained early in the semester (week 3), we orientated classroom activities to likely favor students with transitional or concrete reasoning skills. 
Indeed this raises the need to specifically adapt the first year of our engineering curriculum to progressively improve reasoning skills.

\begin{table}[h]
\caption{\textbf{Average level of Lawson Classroom Test of Scientific Reasoning, and standard deviation as compiled by Mashood \cite{Mashood2019}, and our own measurement (in bold)}}
\begin{ruledtabular}
\begin{tabular}{cccc}
USA & China & India & This study \\
\hline
$74.2\% \pm 18.0 \% $ & $74.7\% \pm 15.8\%$ & $69.3\%  \pm 5.6\%$ & $60.3\% \pm 9.2\%$\\
\end{tabular}
\end{ruledtabular}
\label{Table_Lawson}
\end{table}

\begin{figure}[h]

\includegraphics[width=0.9\linewidth]{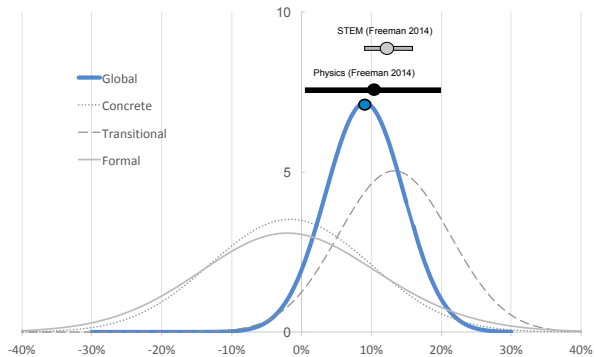}
\caption{\textbf{Difference in failure rate between experimental and control groups}. Probability distributions for the whole student universe as well as distinction of concrete, transitional and formal reasoning skills are shown.}
\label{fig_global}
\end{figure}

\begin{table}[h]
\caption{\textbf{Failure rate in introductory physics:} global result and detail by preinstruction reasoning skills. $p(H_O)$ is the probability that failure rate decreased by introducing active learning. Bold letters are used when $H_O$ is statistically likely.}
\begin{ruledtabular}
\begin{tabular}{cccccc}
& N & traditional & active 
 & variation & $p(H_0)$\\
\hline
Concrete& 73 & 52.6\% & 54.3\% & 1.7\%
& 42.7\% \\
\textbf{Transitional}& \textbf{152} & \textbf{45.0\%} & \textbf{31.5\%} & \textbf{-13.5\%}
& \textbf{94.7\%} \\
Formal& 35 & 18.8\% & 21.1\% & 2.2\%
& 43.2\% \\
\end{tabular}
\end{ruledtabular}
\label{table:fr1}
\end{table}

In  table \ref{table:fr1} we present failure rates separated by reasoning skills for both the experimental and control groups. 
The reasoning skills level is correlated to the probability of passing the course: concrete students were approximately 1.7 less likely to pass the course than transitional students and 2.5 less likely than formal students. 
While we observed a null effect on concrete and formal students, active methodology strongly reduced failure rate of transitional students (-13.5\%) as can be seen in table \ref{table:fr1} and figure \ref{fig_global}. Results are statistically significant.

\section{\label{sec:concl} conclusion}
We implemented evidenced based active learning lectures in an introductory physics course for engineering, supported by a professor training model.
Our study evidences specifically a lower academic risk for students that had not yet developed formal reasoning skills, which are the majority in our local context.

In light of the pilot plan, it is now planed to progressively scale up the  methodology to all sections of introductory physics.
Extrapolation of our findings indicate a potential increment of 150 students approving the course each year. 
In our local context, not exceptional in developing countries, implementing active learning does not increase the number of sections, making it financially sound: the initial investment of one classroom renovation being equivalent to 8 students fees, would be compensated in less than one year by the reduction of academic risk.

\begin{acknowledgments}
We thank to Vicerrectoría Académica USACH through Convenio Marco CM USA1856, and Engineering Faculty of USACH.

C. Hernández is supported by CONICYT through Proyecto Fondecyt de Iniciación No. 11170580.
S. Tecpan, thank to Proyecto DICYT USACH No. 031931TF

The authors would also like to thank the professors Rodrigo Canto, Juan Francisco Fuentealba, Marcia Melendez and the teaching asistant Angel Barra, Ali Godoy, Matias Herrera and Cristobal Hormazabal for their dedication to the project.
\end{acknowledgments}

%\nocite{*}

%\bibliography{PRER_active}% Produces the bibliography via BibTeX.

\begin{thebibliography}{37}%
\makeatletter
\providecommand \@ifxundefined [1]{%
 \@ifx{#1\undefined}
}%
\providecommand \@ifnum [1]{%
 \ifnum #1\expandafter \@firstoftwo
 \else \expandafter \@secondoftwo
 \fi
}%
\providecommand \@ifx [1]{%
 \ifx #1\expandafter \@firstoftwo
 \else \expandafter \@secondoftwo
 \fi
}%
\providecommand \natexlab [1]{#1}%
\providecommand \enquote  [1]{``#1''}%
\providecommand \bibnamefont  [1]{#1}%
\providecommand \bibfnamefont [1]{#1}%
\providecommand \citenamefont [1]{#1}%
\providecommand \href@noop [0]{\@secondoftwo}%
\providecommand \href [0]{\begingroup \@sanitize@url \@href}%
\providecommand \@href[1]{\@@startlink{#1}\@@href}%
\providecommand \@@href[1]{\endgroup#1\@@endlink}%
\providecommand \@sanitize@url [0]{\catcode `\\12\catcode `\$12\catcode
  `\&12\catcode `\#12\catcode `\^12\catcode `\_12\catcode `\%12\relax}%
\providecommand \@@startlink[1]{}%
\providecommand \@@endlink[0]{}%
\providecommand \url  [0]{\begingroup\@sanitize@url \@url }%
\providecommand \@url [1]{\endgroup\@href {#1}{\urlprefix }}%
\providecommand \urlprefix  [0]{URL }%
\providecommand \Eprint [0]{\href }%
\providecommand \doibase [0]{https://doi.org/}%
\providecommand \selectlanguage [0]{\@gobble}%
\providecommand \bibinfo  [0]{\@secondoftwo}%
\providecommand \bibfield  [0]{\@secondoftwo}%
\providecommand \translation [1]{[#1]}%
\providecommand \BibitemOpen [0]{}%
\providecommand \bibitemStop [0]{}%
\providecommand \bibitemNoStop [0]{.\EOS\space}%
\providecommand \EOS [0]{\spacefactor3000\relax}%
\providecommand \BibitemShut  [1]{\csname bibitem#1\endcsname}%
\let\auto@bib@innerbib\@empty
%</preamble>
\bibitem [{\citenamefont {Sithole}\ \emph {et~al.}(2017)\citenamefont
  {Sithole}, \citenamefont {Chiyaka}, \citenamefont {McCarthy}, \citenamefont
  {Mupinga}, \citenamefont {Bucklein},\ and\ \citenamefont
  {Kibirige}}]{sithole2017student}%
  \BibitemOpen
  \bibfield  {author} {\bibinfo {author} {\bibfnamefont {A.}~\bibnamefont
  {Sithole}}, \bibinfo {author} {\bibfnamefont {E.~T.}\ \bibnamefont
  {Chiyaka}}, \bibinfo {author} {\bibfnamefont {P.}~\bibnamefont {McCarthy}},
  \bibinfo {author} {\bibfnamefont {D.~M.}\ \bibnamefont {Mupinga}}, \bibinfo
  {author} {\bibfnamefont {B.~K.}\ \bibnamefont {Bucklein}},\ and\ \bibinfo
  {author} {\bibfnamefont {J.}~\bibnamefont {Kibirige}},\ }\bibfield  {title}
  {\bibinfo {title} {Student attraction, persistence and retention in stem
  programs: Successes and continuing challenges.},\ }\href@noop {} {\bibfield
  {journal} {\bibinfo  {journal} {Higher Education Studies}\ }\textbf {\bibinfo
  {volume} {7}},\ \bibinfo {pages} {46} (\bibinfo {year} {2017})}\BibitemShut
  {NoStop}%
\bibitem [{\citenamefont {Vasquez}\ \emph {et~al.}(2015)\citenamefont
  {Vasquez}, \citenamefont {Fuentes}, \citenamefont {Kypuros},\ and\
  \citenamefont {Azarbayejani}}]{vasquez2015early}%
  \BibitemOpen
  \bibfield  {author} {\bibinfo {author} {\bibfnamefont {H.}~\bibnamefont
  {Vasquez}}, \bibinfo {author} {\bibfnamefont {A.~A.}\ \bibnamefont
  {Fuentes}}, \bibinfo {author} {\bibfnamefont {J.~A.}\ \bibnamefont
  {Kypuros}},\ and\ \bibinfo {author} {\bibfnamefont {M.}~\bibnamefont
  {Azarbayejani}},\ }\bibfield  {title} {\bibinfo {title} {Early identification
  of at-risk students in a lower-level engineering gatekeeper course},\ }in\
  \href@noop {} {\emph {\bibinfo {booktitle} {2015 IEEE Frontiers in Education
  Conference (FIE)}}}\ (\bibinfo {organization} {IEEE},\ \bibinfo {year}
  {2015})\ pp.\ \bibinfo {pages} {1--9}\BibitemShut {NoStop}%
\bibitem [{Note1()}]{Note1}%
  \BibitemOpen
  \bibinfo {note} {MEI usach 2014}\BibitemShut {NoStop}%
\bibitem [{Note2()}]{Note2}%
  \BibitemOpen
  \bibinfo {note} {Https://www.consortium2030.cl/}\BibitemShut {NoStop}%
\bibitem [{\citenamefont {Freeman}\ \emph {et~al.}(2014)\citenamefont
  {Freeman}, \citenamefont {Eddy}, \citenamefont {McDonough}, \citenamefont
  {Smith}, \citenamefont {Okoroafor}, \citenamefont {Jordt},\ and\
  \citenamefont {Wenderoth}}]{freeman2014active}%
  \BibitemOpen
  \bibfield  {author} {\bibinfo {author} {\bibfnamefont {S.}~\bibnamefont
  {Freeman}}, \bibinfo {author} {\bibfnamefont {S.~L.}\ \bibnamefont {Eddy}},
  \bibinfo {author} {\bibfnamefont {M.}~\bibnamefont {McDonough}}, \bibinfo
  {author} {\bibfnamefont {M.~K.}\ \bibnamefont {Smith}}, \bibinfo {author}
  {\bibfnamefont {N.}~\bibnamefont {Okoroafor}}, \bibinfo {author}
  {\bibfnamefont {H.}~\bibnamefont {Jordt}},\ and\ \bibinfo {author}
  {\bibfnamefont {M.~P.}\ \bibnamefont {Wenderoth}},\ }\bibfield  {title}
  {\bibinfo {title} {Active learning increases student performance in science,
  engineering, and mathematics},\ }\href@noop {} {\bibfield  {journal}
  {\bibinfo  {journal} {Proceedings of the National Academy of Sciences}\
  }\textbf {\bibinfo {volume} {111}},\ \bibinfo {pages} {8410} (\bibinfo {year}
  {2014})}\BibitemShut {NoStop}%
\bibitem [{\citenamefont {Crouch}\ and\ \citenamefont
  {Mazur}(2001)}]{crouch2001peer}%
  \BibitemOpen
  \bibfield  {author} {\bibinfo {author} {\bibfnamefont {C.~H.}\ \bibnamefont
  {Crouch}}\ and\ \bibinfo {author} {\bibfnamefont {E.}~\bibnamefont {Mazur}},\
  }\bibfield  {title} {\bibinfo {title} {Peer instruction: Ten years of
  experience and results},\ }\href@noop {} {\bibfield  {journal} {\bibinfo
  {journal} {American journal of physics}\ }\textbf {\bibinfo {volume} {69}},\
  \bibinfo {pages} {970} (\bibinfo {year} {2001})}\BibitemShut {NoStop}%
\bibitem [{\citenamefont {Sharma}\ \emph {et~al.}(2010)\citenamefont {Sharma},
  \citenamefont {Johnston}, \citenamefont {Johnston}, \citenamefont {Varvell},
  \citenamefont {Robertson}, \citenamefont {Hopkins}, \citenamefont {Stewart},
  \citenamefont {Cooper},\ and\ \citenamefont {Thornton}}]{sharma2010use}%
  \BibitemOpen
  \bibfield  {author} {\bibinfo {author} {\bibfnamefont {M.~D.}\ \bibnamefont
  {Sharma}}, \bibinfo {author} {\bibfnamefont {I.~D.}\ \bibnamefont
  {Johnston}}, \bibinfo {author} {\bibfnamefont {H.}~\bibnamefont {Johnston}},
  \bibinfo {author} {\bibfnamefont {K.}~\bibnamefont {Varvell}}, \bibinfo
  {author} {\bibfnamefont {G.}~\bibnamefont {Robertson}}, \bibinfo {author}
  {\bibfnamefont {A.}~\bibnamefont {Hopkins}}, \bibinfo {author} {\bibfnamefont
  {C.}~\bibnamefont {Stewart}}, \bibinfo {author} {\bibfnamefont
  {I.}~\bibnamefont {Cooper}},\ and\ \bibinfo {author} {\bibfnamefont
  {R.}~\bibnamefont {Thornton}},\ }\bibfield  {title} {\bibinfo {title} {Use of
  interactive lecture demonstrations: A ten year study},\ }\href@noop {}
  {\bibfield  {journal} {\bibinfo  {journal} {Physical Review Special
  Topics-Physics Education Research}\ }\textbf {\bibinfo {volume} {6}},\
  \bibinfo {pages} {020119} (\bibinfo {year} {2010})}\BibitemShut {NoStop}%
\bibitem [{\citenamefont {Bubou}\ \emph {et~al.}(2017)\citenamefont {Bubou},
  \citenamefont {Offor},\ and\ \citenamefont {Bappa}}]{bubou2017research}%
  \BibitemOpen
  \bibfield  {author} {\bibinfo {author} {\bibfnamefont {G.~M.}\ \bibnamefont
  {Bubou}}, \bibinfo {author} {\bibfnamefont {I.~T.}\ \bibnamefont {Offor}},\
  and\ \bibinfo {author} {\bibfnamefont {A.~S.}\ \bibnamefont {Bappa}},\
  }\bibfield  {title} {\bibinfo {title} {Why research-informed teaching in
  engineering education? a review of the evidence},\ }\href@noop {} {\bibfield
  {journal} {\bibinfo  {journal} {European Journal of Engineering Education}\
  }\textbf {\bibinfo {volume} {42}},\ \bibinfo {pages} {323} (\bibinfo {year}
  {2017})}\BibitemShut {NoStop}%
\bibitem [{\citenamefont {Santelices}\ \emph {et~al.}(2018)\citenamefont
  {Santelices}, \citenamefont {Catal{\'a}n}, \citenamefont {Horn},\ and\
  \citenamefont {Venegas}}]{santelices2018high}%
  \BibitemOpen
  \bibfield  {author} {\bibinfo {author} {\bibfnamefont {M.~V.}\ \bibnamefont
  {Santelices}}, \bibinfo {author} {\bibfnamefont {X.}~\bibnamefont
  {Catal{\'a}n}}, \bibinfo {author} {\bibfnamefont {C.}~\bibnamefont {Horn}},\
  and\ \bibinfo {author} {\bibfnamefont {A.}~\bibnamefont {Venegas}},\
  }\bibfield  {title} {\bibinfo {title} {High school ranking in university
  admissions at a national level: Theory of action and early results from
  chile},\ }\href@noop {} {\bibfield  {journal} {\bibinfo  {journal} {Higher
  Education Policy}\ }\textbf {\bibinfo {volume} {31}},\ \bibinfo {pages} {159}
  (\bibinfo {year} {2018})}\BibitemShut {NoStop}%
\bibitem [{\citenamefont {Santelices}\ \emph {et~al.}(2019)\citenamefont
  {Santelices}, \citenamefont {Horn},\ and\ \citenamefont
  {Catalán}}]{santelices2019}%
  \BibitemOpen
  \bibfield  {author} {\bibinfo {author} {\bibfnamefont {M.~V.}\ \bibnamefont
  {Santelices}}, \bibinfo {author} {\bibfnamefont {C.}~\bibnamefont {Horn}},\
  and\ \bibinfo {author} {\bibfnamefont {X.}~\bibnamefont {Catalán}},\
  }\bibfield  {title} {\bibinfo {title} {Institution-level admissions
  initiatives in chile: enhancing equity in higher education?},\ }\href
  {https://doi.org/10.1080/03075079.2017.1398722} {\bibfield  {journal}
  {\bibinfo  {journal} {Studies in Higher Education}\ }\textbf {\bibinfo
  {volume} {44}},\ \bibinfo {pages} {733} (\bibinfo {year} {2019})},\ \Eprint
  {https://arxiv.org/abs/https://doi.org/10.1080/03075079.2017.1398722}
  {https://doi.org/10.1080/03075079.2017.1398722} \BibitemShut {NoStop}%
\bibitem [{\citenamefont {Lawson}(2010)}]{lawson2010teaching}%
  \BibitemOpen
  \bibfield  {author} {\bibinfo {author} {\bibfnamefont {A.~E.}\ \bibnamefont
  {Lawson}},\ }\href@noop {} {\emph {\bibinfo {title} {Teaching inquiry science
  in middle and secondary schools}}}\ (\bibinfo  {publisher} {Sage},\ \bibinfo
  {year} {2010})\BibitemShut {NoStop}%
\bibitem [{\citenamefont {Fabby}(2015)}]{fabby2015examining}%
  \BibitemOpen
  \bibfield  {author} {\bibinfo {author} {\bibfnamefont {C.}~\bibnamefont
  {Fabby}},\ }\bibfield  {title} {\bibinfo {title} {Examining the relationship
  of scientific reasoning with physics problem solving},\ }\href@noop {}
  {\bibfield  {journal} {\bibinfo  {journal} {Journal of STEM Education}\
  }\textbf {\bibinfo {volume} {16}} (\bibinfo {year} {2015})}\BibitemShut
  {NoStop}%
\bibitem [{\citenamefont {Coletta}\ \emph {et~al.}(2007)\citenamefont
  {Coletta}, \citenamefont {Phillips},\ and\ \citenamefont
  {Steinert}}]{coletta2007you}%
  \BibitemOpen
  \bibfield  {author} {\bibinfo {author} {\bibfnamefont {V.~P.}\ \bibnamefont
  {Coletta}}, \bibinfo {author} {\bibfnamefont {J.~A.}\ \bibnamefont
  {Phillips}},\ and\ \bibinfo {author} {\bibfnamefont {J.~J.}\ \bibnamefont
  {Steinert}},\ }\bibfield  {title} {\bibinfo {title} {Why you should measure
  your students' reasoning ability},\ }\href@noop {} {\bibfield  {journal}
  {\bibinfo  {journal} {The Physics Teacher}\ }\textbf {\bibinfo {volume}
  {45}},\ \bibinfo {pages} {235} (\bibinfo {year} {2007})}\BibitemShut
  {NoStop}%
\bibitem [{\citenamefont {Coletta}\ and\ \citenamefont
  {Phillips}(2005)}]{coletta2005interpreting}%
  \BibitemOpen
  \bibfield  {author} {\bibinfo {author} {\bibfnamefont {V.~P.}\ \bibnamefont
  {Coletta}}\ and\ \bibinfo {author} {\bibfnamefont {J.~A.}\ \bibnamefont
  {Phillips}},\ }\bibfield  {title} {\bibinfo {title} {Interpreting fci scores:
  Normalized gain, preinstruction scores, and scientific reasoning ability},\
  }\href@noop {} {\bibfield  {journal} {\bibinfo  {journal} {American Journal
  of Physics}\ }\textbf {\bibinfo {volume} {73}},\ \bibinfo {pages} {1172}
  (\bibinfo {year} {2005})}\BibitemShut {NoStop}%
\bibitem [{\citenamefont {Zavala}\ \emph {et~al.}(2007)\citenamefont {Zavala},
  \citenamefont {Alarc{\'o}n},\ and\ \citenamefont
  {Benegas}}]{zavala2007innovative}%
  \BibitemOpen
  \bibfield  {author} {\bibinfo {author} {\bibfnamefont {G.}~\bibnamefont
  {Zavala}}, \bibinfo {author} {\bibfnamefont {H.}~\bibnamefont
  {Alarc{\'o}n}},\ and\ \bibinfo {author} {\bibfnamefont {J.}~\bibnamefont
  {Benegas}},\ }\bibfield  {title} {\bibinfo {title} {Innovative training of
  in-service teachers for active learning: A short teacher development course
  based on physics education research},\ }\href@noop {} {\bibfield  {journal}
  {\bibinfo  {journal} {Journal of Science Teacher Education}\ }\textbf
  {\bibinfo {volume} {18}},\ \bibinfo {pages} {559} (\bibinfo {year}
  {2007})}\BibitemShut {NoStop}%
\bibitem [{\citenamefont {Zavala}\ and\ \citenamefont
  {Alarcon}(2008)}]{zavala2008evaluation}%
  \BibitemOpen
  \bibfield  {author} {\bibinfo {author} {\bibfnamefont {G.}~\bibnamefont
  {Zavala}}\ and\ \bibinfo {author} {\bibfnamefont {H.}~\bibnamefont
  {Alarcon}},\ }\bibfield  {title} {\bibinfo {title} {Evaluation of instruction
  using the conceptual survey of electricity and magnetism in mexico},\ }in\
  \href@noop {} {\emph {\bibinfo {booktitle} {AIP Conference Proceedings}}},\
  Vol.\ \bibinfo {volume} {1064}\ (\bibinfo {organization} {AIP},\ \bibinfo
  {year} {2008})\ pp.\ \bibinfo {pages} {231--234}\BibitemShut {NoStop}%
\bibitem [{\citenamefont {Auyuanet}\ \emph {et~al.}(2018)\citenamefont
  {Auyuanet}, \citenamefont {Modzelewski}, \citenamefont {Loureiro},
  \citenamefont {Alessandrini},\ and\ \citenamefont
  {M{\'\i}guez}}]{auyuanet2018fisicactiva}%
  \BibitemOpen
  \bibfield  {author} {\bibinfo {author} {\bibfnamefont {A.}~\bibnamefont
  {Auyuanet}}, \bibinfo {author} {\bibfnamefont {H.}~\bibnamefont
  {Modzelewski}}, \bibinfo {author} {\bibfnamefont {S.}~\bibnamefont
  {Loureiro}}, \bibinfo {author} {\bibfnamefont {D.}~\bibnamefont
  {Alessandrini}},\ and\ \bibinfo {author} {\bibfnamefont {M.}~\bibnamefont
  {M{\'\i}guez}},\ }\bibfield  {title} {\bibinfo {title} {F{\'\i}sicactiva:
  applying active learning strategies to a large engineering lecture},\
  }\href@noop {} {\bibfield  {journal} {\bibinfo  {journal} {European Journal
  of Engineering Education}\ }\textbf {\bibinfo {volume} {43}},\ \bibinfo
  {pages} {55} (\bibinfo {year} {2018})}\BibitemShut {NoStop}%
\bibitem [{\citenamefont {Hern{\'a}ndez}\ and\ \citenamefont
  {Tecpan}(2018)}]{hernandez2018correct}%
  \BibitemOpen
  \bibfield  {author} {\bibinfo {author} {\bibfnamefont {C.}~\bibnamefont
  {Hern{\'a}ndez}}\ and\ \bibinfo {author} {\bibfnamefont {S.}~\bibnamefont
  {Tecpan}},\ }\bibfield  {title} {\bibinfo {title} {Correct answers with wrong
  justifications? analysis of explanations in classical mechanics with fci
  test},\ }in\ \href@noop {} {\emph {\bibinfo {booktitle} {Journal of Physics:
  Conference Series}}},\ Vol.\ \bibinfo {volume} {1043}\ (\bibinfo
  {organization} {IOP Publishing},\ \bibinfo {year} {2018})\ p.\ \bibinfo
  {pages} {012056}\BibitemShut {NoStop}%
\bibitem [{\citenamefont {Redish}(2014)}]{redish2014oersted}%
  \BibitemOpen
  \bibfield  {author} {\bibinfo {author} {\bibfnamefont {E.~F.}\ \bibnamefont
  {Redish}},\ }\href@noop {} {\bibinfo {title} {Oersted lecture 2013: How
  should we think about how our students think?}} (\bibinfo {year}
  {2014})\BibitemShut {NoStop}%
\bibitem [{\citenamefont {Rudolph}\ \emph {et~al.}(2014)\citenamefont
  {Rudolph}, \citenamefont {Lamine}, \citenamefont {Joyce}, \citenamefont
  {Vignolles},\ and\ \citenamefont {Consiglio}}]{rudolph2014introduction}%
  \BibitemOpen
  \bibfield  {author} {\bibinfo {author} {\bibfnamefont {A.~L.}\ \bibnamefont
  {Rudolph}}, \bibinfo {author} {\bibfnamefont {B.}~\bibnamefont {Lamine}},
  \bibinfo {author} {\bibfnamefont {M.}~\bibnamefont {Joyce}}, \bibinfo
  {author} {\bibfnamefont {H.}~\bibnamefont {Vignolles}},\ and\ \bibinfo
  {author} {\bibfnamefont {D.}~\bibnamefont {Consiglio}},\ }\bibfield  {title}
  {\bibinfo {title} {Introduction of interactive learning into french
  university physics classrooms},\ }\href@noop {} {\bibfield  {journal}
  {\bibinfo  {journal} {Physical review special topics-physics education
  research}\ }\textbf {\bibinfo {volume} {10}},\ \bibinfo {pages} {010103}
  (\bibinfo {year} {2014})}\BibitemShut {NoStop}%
\bibitem [{\citenamefont {Deslauriers}\ \emph {et~al.}(2011)\citenamefont
  {Deslauriers}, \citenamefont {Schelew},\ and\ \citenamefont
  {Wieman}}]{deslauriers2011improved}%
  \BibitemOpen
  \bibfield  {author} {\bibinfo {author} {\bibfnamefont {L.}~\bibnamefont
  {Deslauriers}}, \bibinfo {author} {\bibfnamefont {E.}~\bibnamefont
  {Schelew}},\ and\ \bibinfo {author} {\bibfnamefont {C.}~\bibnamefont
  {Wieman}},\ }\bibfield  {title} {\bibinfo {title} {Improved learning in a
  large-enrollment physics class},\ }\href@noop {} {\bibfield  {journal}
  {\bibinfo  {journal} {science}\ }\textbf {\bibinfo {volume} {332}},\ \bibinfo
  {pages} {862} (\bibinfo {year} {2011})}\BibitemShut {NoStop}%
\bibitem [{\citenamefont {Meltzer}\ and\ \citenamefont
  {Thornton}(2012)}]{meltzer2012resource}%
  \BibitemOpen
  \bibfield  {author} {\bibinfo {author} {\bibfnamefont {D.~E.}\ \bibnamefont
  {Meltzer}}\ and\ \bibinfo {author} {\bibfnamefont {R.~K.}\ \bibnamefont
  {Thornton}},\ }\bibfield  {title} {\bibinfo {title} {Resource letter alip--1:
  active-learning instruction in physics},\ }\href@noop {} {\bibfield
  {journal} {\bibinfo  {journal} {American journal of physics}\ }\textbf
  {\bibinfo {volume} {80}},\ \bibinfo {pages} {478} (\bibinfo {year}
  {2012})}\BibitemShut {NoStop}%
\bibitem [{\citenamefont {Ding}(2013)}]{ding2013detecting}%
  \BibitemOpen
  \bibfield  {author} {\bibinfo {author} {\bibfnamefont {L.}~\bibnamefont
  {Ding}},\ }\bibfield  {title} {\bibinfo {title} {Detecting progression of
  scientific reasoning among university science and engineering students},\
  }in\ \href@noop {} {\emph {\bibinfo {booktitle} {Physics Education Research
  Conference}}}\ (\bibinfo {year} {2013})\ pp.\ \bibinfo {pages}
  {125--128}\BibitemShut {NoStop}%
\bibitem [{\citenamefont {Fabby}\ and\ \citenamefont
  {Koenig}(2013)}]{fabby2013relationship}%
  \BibitemOpen
  \bibfield  {author} {\bibinfo {author} {\bibfnamefont {C.}~\bibnamefont
  {Fabby}}\ and\ \bibinfo {author} {\bibfnamefont {K.}~\bibnamefont {Koenig}},\
  }\bibfield  {title} {\bibinfo {title} {Relationship of scientific reasoning
  to solving different physics problem types},\ }\href@noop {} {\bibfield
  {journal} {\bibinfo  {journal} {Proceedings from PERC Portland, OR, July
  17-18}\ } (\bibinfo {year} {2013})}\BibitemShut {NoStop}%
\bibitem [{\citenamefont {McDermott}\ and\ \citenamefont
  {Shaffer}(1997)}]{mcdermott1997tutoriales}%
  \BibitemOpen
  \bibfield  {author} {\bibinfo {author} {\bibfnamefont {L.~C.}\ \bibnamefont
  {McDermott}}\ and\ \bibinfo {author} {\bibfnamefont {P.~S.}\ \bibnamefont
  {Shaffer}},\ }\href@noop {} {\emph {\bibinfo {title} {Tutoriales para
  f{\'\i}sica introductoria}}}\ (\bibinfo  {publisher} {Pearson
  Educaci{\'o}n},\ \bibinfo {year} {1997})\BibitemShut {NoStop}%
\bibitem [{\citenamefont {Sokoloff}\ and\ \citenamefont
  {Thornton}(2004)}]{sokoloff2004interactive}%
  \BibitemOpen
  \bibfield  {author} {\bibinfo {author} {\bibfnamefont {D.~R.}\ \bibnamefont
  {Sokoloff}}\ and\ \bibinfo {author} {\bibfnamefont {R.~K.}\ \bibnamefont
  {Thornton}},\ }\bibfield  {title} {\bibinfo {title} {Interactive lecture
  demonstrations},\ }\href@noop {} {\bibfield  {journal} {\bibinfo  {journal}
  {Interactive Lecture Demonstrations, by David R. Sokoloff, Ronald K.
  Thornton, pp. 374. ISBN 0-471-48774-0. Wiley-VCH, March 2004.}\ ,\ \bibinfo
  {pages} {374}} (\bibinfo {year} {2004})}\BibitemShut {NoStop}%
\bibitem [{\citenamefont {Mazur}(1999)}]{mazur1999peer}%
  \BibitemOpen
  \bibfield  {author} {\bibinfo {author} {\bibfnamefont {E.}~\bibnamefont
  {Mazur}},\ }\href@noop {} {\bibinfo {title} {Peer instruction: A user’s
  manual}} (\bibinfo {year} {1999})\BibitemShut {NoStop}%
\bibitem [{\citenamefont {Hieggelke}\ \emph {et~al.}(2015)\citenamefont
  {Hieggelke}, \citenamefont {Kanim}, \citenamefont {O'Kuma},\ and\
  \citenamefont {Maloney}}]{hieggelke2015tipers}%
  \BibitemOpen
  \bibfield  {author} {\bibinfo {author} {\bibfnamefont {C.~J.}\ \bibnamefont
  {Hieggelke}}, \bibinfo {author} {\bibfnamefont {S.~E.}\ \bibnamefont
  {Kanim}}, \bibinfo {author} {\bibfnamefont {T.~L.}\ \bibnamefont {O'Kuma}},\
  and\ \bibinfo {author} {\bibfnamefont {D.~P.}\ \bibnamefont {Maloney}},\
  }\href@noop {} {\emph {\bibinfo {title} {TIPERs: Sensemaking Tasks for
  Introductory Physics}}}\ (\bibinfo  {publisher} {Pearson},\ \bibinfo {year}
  {2015})\BibitemShut {NoStop}%
\bibitem [{\citenamefont {Heller}\ \emph {et~al.}(1992)\citenamefont {Heller},
  \citenamefont {Keith},\ and\ \citenamefont {Anderson}}]{heller1992teaching}%
  \BibitemOpen
  \bibfield  {author} {\bibinfo {author} {\bibfnamefont {P.}~\bibnamefont
  {Heller}}, \bibinfo {author} {\bibfnamefont {R.}~\bibnamefont {Keith}},\ and\
  \bibinfo {author} {\bibfnamefont {S.}~\bibnamefont {Anderson}},\ }\bibfield
  {title} {\bibinfo {title} {Teaching problem solving through cooperative
  grouping. part 1: Group versus individual problem solving},\ }\href@noop {}
  {\bibfield  {journal} {\bibinfo  {journal} {American journal of physics}\
  }\textbf {\bibinfo {volume} {60}},\ \bibinfo {pages} {627} (\bibinfo {year}
  {1992})}\BibitemShut {NoStop}%
\bibitem [{\citenamefont {Beichner}\ \emph {et~al.}(2007)\citenamefont
  {Beichner}, \citenamefont {Saul}, \citenamefont {Abbott}, \citenamefont
  {Morse}, \citenamefont {Deardorff}, \citenamefont {Allain}, \citenamefont
  {Bonham}, \citenamefont {Dancy},\ and\ \citenamefont
  {Risley}}]{beichner2007student}%
  \BibitemOpen
  \bibfield  {author} {\bibinfo {author} {\bibfnamefont {R.~J.}\ \bibnamefont
  {Beichner}}, \bibinfo {author} {\bibfnamefont {J.~M.}\ \bibnamefont {Saul}},
  \bibinfo {author} {\bibfnamefont {D.~S.}\ \bibnamefont {Abbott}}, \bibinfo
  {author} {\bibfnamefont {J.~J.}\ \bibnamefont {Morse}}, \bibinfo {author}
  {\bibfnamefont {D.}~\bibnamefont {Deardorff}}, \bibinfo {author}
  {\bibfnamefont {R.~J.}\ \bibnamefont {Allain}}, \bibinfo {author}
  {\bibfnamefont {S.~W.}\ \bibnamefont {Bonham}}, \bibinfo {author}
  {\bibfnamefont {M.~H.}\ \bibnamefont {Dancy}},\ and\ \bibinfo {author}
  {\bibfnamefont {J.~S.}\ \bibnamefont {Risley}},\ }\bibfield  {title}
  {\bibinfo {title} {The student-centered activities for large enrollment
  undergraduate programs (scale-up) project},\ }\href@noop {} {\bibfield
  {journal} {\bibinfo  {journal} {Research-based reform of university physics}\
  }\textbf {\bibinfo {volume} {1}},\ \bibinfo {pages} {2} (\bibinfo {year}
  {2007})}\BibitemShut {NoStop}%
\bibitem [{\citenamefont {Fox-Turnbull}\ \emph {et~al.}(2018)\citenamefont
  {Fox-Turnbull}, \citenamefont {Docherty},\ and\ \citenamefont
  {Zaka}}]{fox2018learning}%
  \BibitemOpen
  \bibfield  {author} {\bibinfo {author} {\bibfnamefont {W.~H.}\ \bibnamefont
  {Fox-Turnbull}}, \bibinfo {author} {\bibfnamefont {P.}~\bibnamefont
  {Docherty}},\ and\ \bibinfo {author} {\bibfnamefont {P.}~\bibnamefont
  {Zaka}},\ }\bibfield  {title} {\bibinfo {title} {Learning engineering through
  the flipped classroom approach-students' perspectives},\ }\href@noop {}
  {\bibfield  {journal} {\bibinfo  {journal} {Design and Technology Education:
  an International Journal}\ }\textbf {\bibinfo {volume} {23}},\ \bibinfo
  {pages} {26} (\bibinfo {year} {2018})}\BibitemShut {NoStop}%
\bibitem [{\citenamefont {Hern{\'a}ndez-Silva}\ and\ \citenamefont
  {Tecpan~Flores}(2017)}]{hernandez2017aula}%
  \BibitemOpen
  \bibfield  {author} {\bibinfo {author} {\bibfnamefont {C.}~\bibnamefont
  {Hern{\'a}ndez-Silva}}\ and\ \bibinfo {author} {\bibfnamefont
  {S.}~\bibnamefont {Tecpan~Flores}},\ }\bibfield  {title} {\bibinfo {title}
  {Aula invertida mediada por el uso de plataformas virtuales: un estudio de
  caso en la formaci{\'o}n de profesores de f{\'\i}sica},\ }\href@noop {}
  {\bibfield  {journal} {\bibinfo  {journal} {Estudios pedag{\'o}gicos
  (Valdivia)}\ }\textbf {\bibinfo {volume} {43}},\ \bibinfo {pages} {193}
  (\bibinfo {year} {2017})}\BibitemShut {NoStop}%
\bibitem [{\citenamefont {Brewe}\ \emph {et~al.}(2018)\citenamefont {Brewe},
  \citenamefont {Dou},\ and\ \citenamefont {Shand}}]{brewe2018costs}%
  \BibitemOpen
  \bibfield  {author} {\bibinfo {author} {\bibfnamefont {E.}~\bibnamefont
  {Brewe}}, \bibinfo {author} {\bibfnamefont {R.}~\bibnamefont {Dou}},\ and\
  \bibinfo {author} {\bibfnamefont {R.}~\bibnamefont {Shand}},\ }\bibfield
  {title} {\bibinfo {title} {Costs of success: Financial implications of
  implementation of active learning in introductory physics courses for
  students and administrators},\ }\href@noop {} {\bibfield  {journal} {\bibinfo
   {journal} {Physical Review Physics Education Research}\ }\textbf {\bibinfo
  {volume} {14}},\ \bibinfo {pages} {010109} (\bibinfo {year}
  {2018})}\BibitemShut {NoStop}%
\bibitem [{\citenamefont {Bao}\ \emph {et~al.}(2018)\citenamefont {Bao},
  \citenamefont {Xiao}, \citenamefont {Koenig},\ and\ \citenamefont
  {Han}}]{Bao2018}%
  \BibitemOpen
  \bibfield  {author} {\bibinfo {author} {\bibfnamefont {L.}~\bibnamefont
  {Bao}}, \bibinfo {author} {\bibfnamefont {Y.}~\bibnamefont {Xiao}}, \bibinfo
  {author} {\bibfnamefont {K.}~\bibnamefont {Koenig}},\ and\ \bibinfo {author}
  {\bibfnamefont {J.}~\bibnamefont {Han}},\ }\bibfield  {title} {\bibinfo
  {title} {Validity evaluation of the lawson classroom test of scientific
  reasoning},\ }\href@noop {} {\bibfield  {journal} {\bibinfo  {journal}
  {Physical Review Physics Education Research}\ }\textbf {\bibinfo {volume}
  {14}},\ \bibinfo {pages} {020106} (\bibinfo {year} {2018})}\BibitemShut
  {NoStop}%
\bibitem [{\citenamefont {Paterno}(2004)}]{Paterno2004}%
  \BibitemOpen
  \bibfield  {author} {\bibinfo {author} {\bibfnamefont {M.}~\bibnamefont
  {Paterno}},\ }\href@noop {} {\emph {\bibinfo {title} {Calculating
  efficiencies and their uncertainties}}},\ \bibinfo {type} {Tech. Rep.}\
  (\bibinfo  {institution} {Fermi National Accelerator Lab.(FNAL), Batavia, IL
  (United States)},\ \bibinfo {year} {2004})\BibitemShut {NoStop}%
\bibitem [{\citenamefont {Bao}\ \emph {et~al.}(2009)\citenamefont {Bao},
  \citenamefont {Cai}, \citenamefont {Koenig}, \citenamefont {Fang},
  \citenamefont {Han}, \citenamefont {Wang}, \citenamefont {Liu}, \citenamefont
  {Ding}, \citenamefont {Cui}, \citenamefont {Luo} \emph {et~al.}}]{Bao2009}%
  \BibitemOpen
  \bibfield  {author} {\bibinfo {author} {\bibfnamefont {L.}~\bibnamefont
  {Bao}}, \bibinfo {author} {\bibfnamefont {T.}~\bibnamefont {Cai}}, \bibinfo
  {author} {\bibfnamefont {K.}~\bibnamefont {Koenig}}, \bibinfo {author}
  {\bibfnamefont {K.}~\bibnamefont {Fang}}, \bibinfo {author} {\bibfnamefont
  {J.}~\bibnamefont {Han}}, \bibinfo {author} {\bibfnamefont {J.}~\bibnamefont
  {Wang}}, \bibinfo {author} {\bibfnamefont {Q.}~\bibnamefont {Liu}}, \bibinfo
  {author} {\bibfnamefont {L.}~\bibnamefont {Ding}}, \bibinfo {author}
  {\bibfnamefont {L.}~\bibnamefont {Cui}}, \bibinfo {author} {\bibfnamefont
  {Y.}~\bibnamefont {Luo}}, \emph {et~al.},\ }\bibfield  {title} {\bibinfo
  {title} {Learning and scientific reasoning},\ }\href@noop {} {\bibfield
  {journal} {\bibinfo  {journal} {Science}\ }\textbf {\bibinfo {volume}
  {323}},\ \bibinfo {pages} {586} (\bibinfo {year} {2009})}\BibitemShut
  {NoStop}%
\bibitem [{\citenamefont {Mashood}\ and\ \citenamefont
  {Singh}(2019)}]{Mashood2019}%
  \BibitemOpen
  \bibfield  {author} {\bibinfo {author} {\bibfnamefont {K.}~\bibnamefont
  {Mashood}}\ and\ \bibinfo {author} {\bibfnamefont {V.~A.}\ \bibnamefont
  {Singh}},\ }\bibfield  {title} {\bibinfo {title} {Preuniversity science
  education in india: Insights and cross cultural comparison},\ }\href@noop {}
  {\bibfield  {journal} {\bibinfo  {journal} {Physical Review Physics Education
  Research}\ }\textbf {\bibinfo {volume} {15}},\ \bibinfo {pages} {013103}
  (\bibinfo {year} {2019})}\BibitemShut {NoStop}%
\end{thebibliography}
%biblio para someter

\end{document}

\preprint{APS/123-QED}

\title{Active Learning reduce academic risk of students with non-formal reasoning skills.\\ Evidence from an introductory physics massive course in a Chilean public university. Supplementary material}

\author{Guillaume Lagubeau}
\email{guillaume.lagubeau@usach.cl}%Lines break automatically or can be forced with \\

\author{Silvia Tecpan}
\email{silvia.tecpan@usach.cl}
\author{Carla Hernandez}
\email{carla.hernandez.s@usach.cl}
 
\affiliation{%
 Universidad de Santiago de Chile, Departamento de Física
}%

\date{\today}% It is always \today, today,
             %  but any date may be explicitly specified

\keywords{Suggested keywords}%Use showkeys class option if keyword
                              %display desired
%\maketitle

\onecolumngrid
\appendix
\section*{Supplementary Material}

In Table \ref{Table:content} we reference the learning activities in chronological order. Evaluations were taken after themes : 
\begin{itemize}
\item Units and measure, Vectors and Forces
\item Torque and structure 
\item Distributed Load and Hydrostatics
\end{itemize}

\begin{table*}[h]
\caption{\textbf{List of learning activities, in chronological order}}
\begin{ruledtabular}
\begin{tabular}{cccc}
Theme & Learning resource & Reference
\\
\hline
Units and measures & Classroom Lawson Test of Scientific Reasoning & \cite{lawson2010teaching}  
\\
Units and measures & PEER instruction:  orders of magnitude  & Own creation 
\\
Units and measures & Collaborative Solving Problems: ``Purchase order"  & Own creation 
\\
Units and measures & Worksheet: classification of scalar and vector quantities  & Own creation 
\\
Vectors &  Sense making tasks: Classify and identify vectors. A2-QRT01 & \cite{hieggelke2015tipers} 
\\
Vectors &  PEER instruction : unit vector & \cite{barniol2014students} 
\\
Vectors &  Tutorial: Vector components & \cite{barniol2015calculation} 
\\
Force &  Tutorial: force diagrams  & Own creation 
\\
Force &  Tutorial: force equilibrium & \cite{mcdermott1997tutoriales}
\\
Force &  PEER instruction: action and reaction pair & Own creation
\\
Force &  Collaborative solving problems: ``hanging sculpture" & Own creation, based on \cite{serway2005fisica}
\\ 
Torque & Balancing Act: PhET simulation, act‬ 1.1.16 & \cite{lehtinen2017guidance}
\\
Torque & Tutorial: torque & \cite{van2006physics}
\\
Torque & Collaborative Solving Problems: firefighter & Own creation, based on \cite{serway2005fisica}
\\
Torque & Collaborative Solving Problems: flexed person & \cite{van2006physics}
\\
Torque & Collaborative Solving Problems: suspended beam & \cite{van2006physics}
\\
Structure & Video: ``types of structures" and discussion &
\footnote{https://www.youtube.com/watch?v=VnBel53lrag}
\\
Structure & Tutorial: structure elements &  Own creation based on \cite{beer2016statics}
\\
Structure &   Video: ``tipos de Apoyos" and discussion & \footnote{https://www.youtube.com/watch?v=_Xf4s0terv4}
\\
Structure &   Tutorial: structure equilibrium &   Own creation based on \cite{beer2016statics}
\\ 
Distributed Load & Video: ``puente colgante soporta pesada carga" and discussion & \footnote{https://www.youtube.com/watch?v=hApCZ0iE6RU}
\\
Distributed Load & Tutorial: Distributed Load & Own creation, based on \cite{beer2016statics}

\\
Distributed Load & Peer instruction: Distributed Load & Own creation, based on \cite{beer2016statics}
\\
Hydrostatics & Sensmaking tasks: C2 Fluids, C2-RT01, C2-RT03& \cite{hieggelke2015tipers} 
\\
Hydrostatics & Sensmaking tasks: C2 Fluids,  C2-RT05,  C2-RT09& \cite{hieggelke2015tipers} 
\\
Hydrostatics & Sensmaking tasks, C2 Fluids,  C2-CT11, C2-CT12 & \cite{hieggelke2015tipers} 

\end{tabular}
\end{ruledtabular}
\label{Table:content}
\end{table*}

%\nocite{*}

%\bibliography{PRER_active}% Produces the bibliography via BibTeX.
%biblio para someter